\documentclass{mn2e}
\usepackage{psfig}
\usepackage{epsf}
\newcommand{\ltaraw}{$\; \buildrel < \over \sim \;$}
\newcommand{\lta}{\lower.5ex\hbox{\ltaraw}}
\newcommand{\gtaraw}{$\; \buildrel > \over \sim \;$}
\newcommand{\gta}{\lower.5ex\hbox{\gtaraw}}

\newcommand{\ffffff}[1]{\mbox{$#1$}}
\newcommand{\scnd}{\mbox{\ffffff{''}\hskip-0.3em.}}

\newcommand{\apm}{APM08279+5255}

\loadboldmathitalic 
\title [STIS spectra of \apm]
{Spatially resolved STIS spectra of the gravitationally 
lensed BAL quasar \apm: the nature of component C
and evidence for microlensing}
\author[G. F. Lewis et al.]
{Geraint F. Lewis$^{1}$, Rodrigo A. Ibata$^{2}$, Sara L. Ellison$^{3}$,
Bastien Aracil$^{4}$, \newauthor 
Patrick Petitjean$^{5}$, Max Pettini$^{6}$, Raghunathan Srianand$^{7}$ \\ 
$^{1}$
Anglo-Australian Observatory, P.O. Box 296, Epping, NSW 1710, Australia:
Email \tt{gfl@aaoepp.aao.gov.au}\\
$^{2}$
Observatoire de Strabourg, 11, rue de l'Universite, F-67000, Strasbourg, 
France:
Email \tt{ibata@pleiades.u-strasbg.fr} \\
$^{3}$
European Southern Observatory, Casilla 19001, Santiago 19, Chile:
Email \tt{sellison@eso.org} \\
$^{4}$
Institut d'Astrophysique de Paris -- CNRS, 98bis Boulevard 
   Arago, F-75014 Paris, France:
Email \tt{aracil@iap.fr} \\
$^{5}$
Institut d'Astrophysique de Paris -- CNRS, 98bis Boulevard 
   Arago, F-75014 Paris, France:
Email \tt{petitjean@iap.fr} \\
$^{6}$
Institute of Astronomy, Madingley Rd, Cambridge, CB3 0HA, U.K.
Email \tt{pettini@ast.cam.ac.uk}\\
$^{7}$
IUCAA, Post Bag 4, Ganeshkhind, Pune 411 007, India:
Email \tt{anand@iucaa.ernet.in}
}
\date{\today}
\begin{document} 
\maketitle 
\begin{abstract}
While gravitationally  lensed quasars are  expected to display  an odd
number of images, invariably systems  are observed with an even number
of  quasars.  For  this, lensing  galaxies must  have very  small core
radii;  this provides  strong demagnification  of one  of  the images.
High  resolution imaging  of  the gravitationally  lensed BAL  quasar,
\apm, reveals three point-like images. As these images possess similar
colours, it  has been suggested  that each represents a  lensed image.
Here,  spatially  resolved   spectra  of  the  individual  components,
obtained with STIS  on the HST, are presented,  clearly revealing that
each is  an image of the  quasar. This confirms  that \apm\ represents
the first example of  an odd-image gravitationally lensed system.  The
implications for  the properties of the lensing  galaxy are discussed.
It  is  also  found   that  the  individual  images  possess  spectral
differences indicative of  the influence of gravitational microlensing
in this system.
\end{abstract}

\section{Introduction}\label{introduction}
The $z=3.911$ broad absorption  line (BAL) quasar \apm\ was identified
serendipitously within a  survey of carbon stars in  the Galactic halo
(Irwin  et al.   1998).  The  optical  emission is  coincident with  a
$\sim1$Jy  IRAS  source  at 100$\mu$m  and  was  also  found to  be  a
significant submillimetre  source, with a  flux of 75mJy  at 850$\mu$m
(Lewis  et   al.   1998);   the  inferred  bolometric   luminosity  is
$\sim5\times10^{15}L_\odot$,  making \apm\  one of  the  most luminous
sources  known.  The  discovery images  revealed that  \apm\ is  not a
single point-like source, rather it  is extended (Irwin et al.  1998).
Adaptive  optics  with  the  CFHT clearly  displayed  \apm's  compound
nature, revealing a pair of  point-like images separated by 0.4 arcsec
(Ledoux et al. 1998). Further observations, using NICMOS on the Hubble
Space Telescope (Ibata et al.   1999) and the Keck telescope (Egami et
al. 2000), uncovered  a fainter third image between  the brighter two,
the colours of  which suggest that it represents a  third image of the
quasar.

Paradoxically, this  conclusion is somewhat  problematic. Although one
of the  fundamental predictions of  gravitational lens theory  is that
there  should always be  an odd  number of  lensed images  (e.g. Burke
1981),  in practice  all lensed  QSOs known  to date  exhibit  an even
number of images.  Narasimha,  Subramanian, \& Citre (1986) have shown
that this is in fact expected  if the lensing galaxies have very small
core  radii;  the core  `captures'  one  of  the images  and  strongly
demagnifies it.   The fact that APM~08279+5255  apparently defies this
trend  is reflected  in the  mass models  for the  lens  which require
unphysically large cores to explain  the brightness of image C, if the
latter is  indeed an  image of the  QSO (Ibata  et al. 1999;  Egami et
al. 2000; Munoz  et al. 2001).  This led Ibata et  al.  (1999) to also
propose an alternative model where component C is actually the lensing
galaxy  responsible   for  the  observed   configuration;  this  model
possesses  more `typical' galaxy  parameters.  Recent  observations of
the  nuclear CO(1-0) emission  in \apm\  reveals a  complex morphology
which  suggests that the  lens in  this system  is a  highly flattened
system, such as  an edge-on spiral galaxy (Lewis  et al.  2002).  With
this model,  component C  is an  image of the  quasar and  the ternary
configuration of the  quasar source lying in the  vicinity of a `naked
cusp' (e.g.  Bartelmann \& Loeb 1998).

\begin{figure}
\centerline{ \psfig{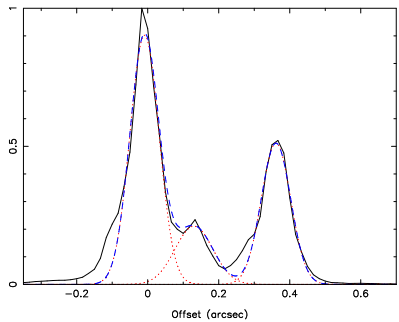} }
\caption{Flux  distribution   along  the  spatial   direction  of  the
slit.  From left  to right  the peaks  correspond to  images A,  C and
B. The three  component Gaussian fit is displayed  with a dotted line,
and the dashed line shows the sum of the three Gaussians.}
\label{Figure1}
\end{figure}

The situation is  complicated by the presence of  two very strong MgII
systems at z=1.062  and 1.181.  A detailed study  of the corresponding
absorption spectrum  shows that these  systems are damped with  a very
high  HI column density  (${\rm >10^{21}  cm^{-2}}$, Petitjean  et al.
2000).   The objects  associated  with these  two  systems could  both
contribute to the lensing of the quasar.

Given the extreme apparent properties  of \apm, an accurate lens model
is   essential   for  the   determination   of   its  true   intrinsic
properties. This paper presents  new spatially resolved spectra of the
various  components   in  this  lensing  system,  with   the  goal  of
determining the nature  of component C.  Section~\ref{obsred} presents
a  description of  the  observing and  data  reduction, reviewing  the
nature  of component  C in  Section~\ref{discuss}.  This  section also
discusses the  implications of  this study.  Finally,  the conclusions
are presented in Section~\ref{conclusions}.

\section{Observations and Reduction}\label{obsred}

\subsection{Observations}\label{obs}
The aim of  the observing program was to  obtain a spatially resolved,
high resolution spectrum of  the $z=3.911$ BAL quasar, APM~08279+5255.
While being  triply imaged, the  image separations in this  system are
small,  ranging  from  0.15~arcsecs  to  a  maximum  of  0.38~arcsecs.
Ground-based observations  of this bright source have  revealed a rich
absorption  spectrum  caused  by  both intervening  material  and  the
complex QSO environment (Ellison et al. 1999a,b; Srianand \& Petitjean
2000; Petitjean et al. 2000).

The Space  Telescope Imaging  Spectrograph (STIS) on-board  the Hubble
Space Telescope  (HST) was employed  in this endeavor.   The principal
aim was to  probe the numerous intervening systems  on sub-kpc scales,
to  investigate the structure  of intervening  galaxy halos  and metal
line systems on  scales of $\sim$ 0.2 -- 1.6 kpc  h$^{-1}$, as well as
their kinematics  and spatial extents.   A further goal was  to obtain
multiple  sightlines through the  complex BAL  flow on  parsec scales,
yielding information  on ionization, kinematics  and metal enrichment;
these will be the topics  of forthcoming articles.  Here we present an
initial investigation of the  spectra in the individual lensed images,
with the goal of establishing the nature of image C.

The STIS G750M grating was used in five different (primary) wavelength
settings so as  to achieve complete coverage from  $\sim 6000$ -- 8600
\AA\  (see Table  1).  Acquisition  was achieved  using  the brightest
component A. The  52 $\times$ 0.2 arcsec slit  was then oriented along
the major axis between the two brightest components (A and B) and then
offset perpendicularly by 0.02 arcseconds to achieve improved centring
on  the faint  component C.   The STIS  spatial pixel  scale  is 0.051
arcsecs/pixel,  with  a   typical  dispersion  of  0.56\AA/pixel.   In
addition to the standard  calibrations (including spectra of the He-Ar
arc  lamp for wavelength  calibration), extra  flat field  images were
obtained  for the  three  reddest  settings in  order  to correct  for
significant  fringing at these  wavelengths.  Multiple  exposures were
obtained for each setting, subsequent exposures stepped along the slit
by 1 arcsecond.

Due to  a combination  of narrow scheduling  windows and  backlog from
previous  cycles,  only  a  subset  of  the orbits  has  so  far  been
completed, see Table 1. While we defer a comprehensive analysis of the
absorption spectra until the full data set has been acquired, there is
sufficient   information   in  the   first   wavelength  setting,   at
$\lambda_{\rm  central}  = 6252$\,\AA\  (first  line  of  Table 1)  to
confirm the nature of component  C.  The data considered in this paper
consist of five individual  spectroscopic exposures; each exposure was
offset by 20  pixels (one arcsecond) relative to  the previous one, so
as to maintain optimal spatial  resolution and minimise the effects of
CCD and camera artifacts.

\begin{table}
\begin{center}
\caption{Summary of Observations}
\begin{tabular}{lrr} \hline \hline
Wavelength &Total &
Completed \\ 
 Range (\AA)  & Integration (s) &
Integration (s) \\ \hline
5965 -- 6538 & 14 900 & 14 900\\
6482 -- 7054 & 11 800 & 0\\
6997 -- 7569 & 11 800 & 8 700\\
7509 -- 8081 & 14 900 & 14 900\\
8025 -- 8597 & 21 100 & 6 200 \\ \hline
\end{tabular}
\end{center}
\end{table}

\subsection{Data Reduction}\label{red}
The extraction  of the individual  spectra is slightly  complicated by
the  fact that the  spectra are  not completely  separated, especially
components A and C, even  with the superb resolution afforded by STIS.
The extraction  was performed  in a straightforward,  but non-standard
way, which  we now detail.  The  adopted procedure is  similar to that
undertaken for  the complex gravitational lens Q2237+0305  by Lewis et
al. (1998).

The first  step we took  was to trace  the spectrum of of  component A
(the brightest QSO component) with a straight line fit; this gives the
spatial  offset between  the  spectral images  to $\sim  0.02$~pixels.
After correcting for the slope of the spectrum on the image, we obtain
the distribution of flux in the spatial direction, integrated over the
wavelength  range, $5965$\AA\  to $6534$\AA  (Figure 1).   To  aid the
visual  interpretation  of this  diagram,  we  have  overlaid a  three
component  Gaussian  fit.  The  three  quasar  components are  clearly
distinguished in  this single spectroscopic exposure;  also evident is
the peculiar asymmetric  profile (seen as an excess  over the Gaussian
fit to the left side of components A and B).

We  constructed  a  model  point-spread  function  using  an  archival
spectrum  of  the star  51~Peg  (HST  root  name O6IH50050)  taken  in
approximately  the same  wavelength range,  only  a day  prior to  the
observations  described here.   By fitting  this  PSF to  the data  in
Figure~1, we are able to define suitable extraction bands for each one
of the components.  These bands  are listed in Table~2, with positions
relative   to  the   centre  of   component  A.    Estimates   of  the
cross-contamination from the other  quasar components and the fraction
of the flux missed are also given in the table (assuming the model PSF
derived from the 51~Peg observation).  The worst case is clearly image
C,  where  the   extracted  spectrum  has  a  total   of  $\sim  10$\%
contamination from components  A and B, and is  missing $\sim 30$\% of
the flux.

\begin{table}
\begin{center}
\caption{Summary of Data Reduction}
\begin{tabular}{lrrrr} \hline \hline
Image & Lower & Upper &
cross- & flux \\ 
      & band edge & band edge &
contamination & missed \\ \hline
A & $-0\scnd15$  & $0\scnd05$ &  2\% &  7\% \\
B &  $0\scnd25$  & $0\scnd50$ &  1\% &  1\% \\
C &  $0\scnd10$  & $0\scnd20$ & 10\% & 30\% \\ \hline
\end{tabular}
\end{center}
\end{table}

He-Ar arc-lamp spectra were  also extracted from the arc-lamp spectral
image from  identical regions as  the object spectra.   These arc-lamp
spectra  were used to  wavelength calibrate  the observed  QSO spectra
which  were  then rebinned  onto  the  interval $\lambda=5965$\AA\  to
$\lambda=6534$\AA\  with linear wavelength  steps.  Finally,  the five
spectral observations  of the three  QSO images were combined  using a
median-combining  algorithm.   Figure~2  displays  the  three  spectra
obtained in this manner.

\begin{figure}
\centerline{ \psfig{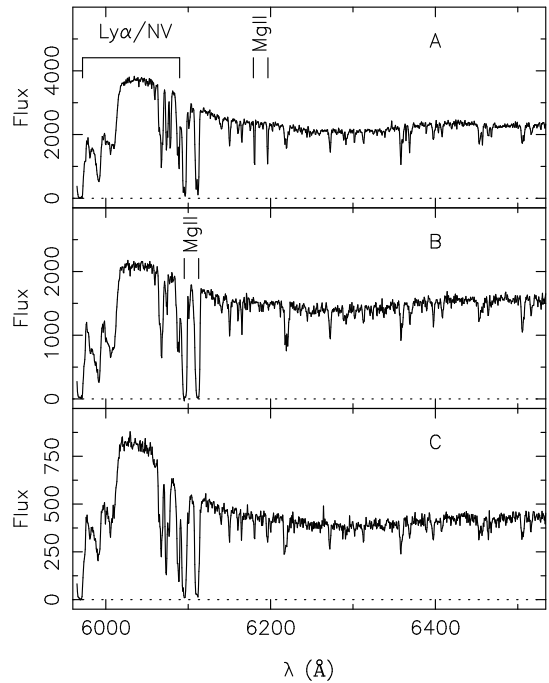} }
\caption{The spectra of  the three QSO images A, B and  C (from top to
bottom).  While  differences in  the individual spectra  are apparent,
the lower  panel confirms that image C  is of the quasar.   As well as
the prominent  emission line  structure due to  ${\rm Ly_\alpha/N~V}$,
strong MgII  $\lambda\lambda 2796,2803$ absorption at z=1.18  due to a
foreground galaxy  is labelled.  One striking feature  is the presence
of MgII  absorption at  z=1.21 ($\sim6190$\AA), also  labelled, which,
while being  strong in  image A,  is weaker in  C and  non-existent in
image B,  suggesting a non-uniform  covering factor across  the images
(Petitjean et al. 2000).
\label{spectra}}
\label{Figure2}
\end{figure}

\section{Discussion}\label{discuss}
\subsection{The nature of component C}\label{compc}
A cursory  examination of the  individual spectra of the  three images
presented in Figure~\ref{spectra} clearly  reveals that each is of the
quasar source, and, therefore, image C represents a third image of the
quasar  and  is  not   the  lensing  galaxy.   This  confirms  earlier
interpretations of the available  photometric data (Ibata et al. 1999;
Egami et al. 2000; Munoz et al. 2001).

\subsection{Spectral differences}\label{diffs}
A more detailed examination of the three spectra in Figure 2, however,
also reveals  that they  are not identical.   Even accounting  for the
strong Mg~II doublet lines near 6100\,\AA\ (at $z_{\rm abs} = 1.181$),
probably  due to  the  presence of  the  lensing galaxy,  and for  the
Ly$\alpha$ and N~V BAL features  below $\sim 6040$\,\AA, it is evident
that the  equivalent width of  the Ly$\alpha$+N~V emission  line blend
near 6050\,\AA\ is different in each of the three images.  Relative to
the continuum, the emission lines are strongest in image C and weakest
in B.

These differences are  summarised in Table 3 (where  we have corrected
for the  missing flux in  the spectrum extraction procedure).  For the
continuum level  we adopted the  mean observed flux in  the wavelength
interval 6250--6350\,\AA\ which, at  $z_{\rm em} = 3.911$, corresponds
to  rest  wavelengths  1273--1293\,\AA;  this  is  a  region  free  of
prominent  emission lines  and  provides  a good  measure  of the  QSO
continuum (Francis  et al.   1991).  The peak  emission line  flux was
measured  between 6040\,\AA\ and  6060\,\AA\ (1230--1234\,\AA\  in the
rest  frame  of   APM~08279+5255)  after  subtracting  the  underlying
continuum.  It can  be seen from Table 3 that the  contrast of image C
relative to  A and B is more  pronounced in the continuum  than in the
emission lines; when viewed in the emission line light, images B and C
are of comparable brightness.

\begin{table}
\begin{center}
\caption{Continuum to emission line flux ratios}
\begin{tabular}{lcc} \hline \hline
Image & Flux (continuum) &
Flux (Ly$_\alpha$ \& N~V) \\ \hline
A & 1.00 &  1.00 \\
B & 0.63 &  0.40 \\
C & 0.23 &  0.38 \\ \hline
\end{tabular}
\end{center}
\end{table}

Such spectral  differences are  a natural consequence  of differential
microlensing effects,  with the small continuum  emitting source being
more severely influenced by the action of microlensing than the larger
line  emitting region  (e.g. Saust  1994).  In  fact, it  is generally
thought that the scale of the broad emission line region is sufficient
for it to be immune from significant microlensing influences (Nemiroff
1988;  Schneider  \&  Wambsganss  1990),  and hence  it  reflects  the
magnification  due  to the  macrolens.   While unresolved  photometric
monitoring  of APM08279+5255  reveals that  this system  has exhibited
pronounced variability over several tenths of a magnitude (Lewis, Robb
\& Ibata 1999; see also  the continuing monitoring program at the Wise
Observatory  at  {\tt http://wise-obs.tau.ac.il/$\sim$eran/LM/}),  the
spectroscopic evidence  presented here  points to the  variation being
potentially due to gravitational microlensing.

Hence, we conclude that the macrolensing magnifications of images B \&
C, based on  the emission lines, are comparable, and  it is these that
should be used  in gravitational lens modeling of  this system, rather
than the continuum flux. We note that a similar conclusion was reached
by  Lewis et  al (2002)  who, based  on extended  CO emission  in this
system,  demonstrated  that the  likely  lens  is  a highly  flattened
object, such  as an edge-on  spiral. While their model  predicted that
these images should  be $\sim75\%$ the brightness of  image A, they do
concede that  currently there are  not enough constraints  to uniquely
tie-down  a  model.   The  magnifications  presented  in  this  paper,
therefore,  will   aid  in  the   modeling  of  this   unique  ternary
gravitational lens system. It is important to note, however, that dust
in the lensing galaxy may result  in extinction of some of the quasars
flux, especially  image C which  may be viewed  through the disk  of a
spiral  system, and  the values  presented  in Table~3  may not  truly
reflect the magnifications of  the macromodel.  While dust can extinct
the  quasar  images,  it  cannot  be  responsible  for  the  differing
line-to-continuum ratios  observed in the STIS spectra,  and hence the
conclusion of gravitational microlensing in this system is robust.

It is  also interesting  to note that,  compared to the  emission line
flux,  while the  continuum in  image B  appears to  be  enhanced, the
continuum in image C appears  to be depressed. Such a situation occurs
during  a microlensing  demagnification  and is  seen dramatically  in
image D  of the quadruple lens  Q2237+0305 (Lewis et  al.  1998). With
such a delineation in microlensing effects, an estimate of the size of
the emission  regions can be made. Regions  significantly smaller than
the  gravitational  microlensing  scale-length, the  Einstein  radius,
(i.e.  the continuum emitting  region) can be significantly influenced
during  microlensing, while  regions significantly  larger  (e.g.  the
broad  emission  line region)  are  not.   The  source plane  Einstein
radius, $\eta$, is given by
\begin{equation}
\eta =  \sqrt{ \frac{4 G M}{c^2} \frac{D_{os}  D_{ls}}{D_{ol}} } 
\end{equation}
where $D_{ij}$ are the  angular diameter distances between an observer
(o),  lens   (l)  and   source  (s).   For   \apm,  $\eta   \sim  0.01
\sqrt{M/M_\odot}  h_{50}^{-\frac{1}{2}}{\rm pc}$  for  an $\Omega_o=1$
cosmology,  where $M$ is  the typical  microlensing mass.   Of course,
confirmation  of  the  gravitational  lens nature  of  these  spectral
features,   and   hence   the   applicability  of   this   delineation
scale-length,  requires  further  time resolved  spectroscopy.   While
spatially resolved spectroscopy, as  presented in this paper, would be
ideal,  variability in  the equivalent  widths of  the  emission lines
should be apparent in unresolved ground-based spectra.

\subsection{Where is the lensing galaxy?}\label{wheregalaxy}
The spectra were  examined for a signature of  the lensing galaxy. The
models  of Ibata  et al.  (1999),  Egami et  al. (2000)  and Munoz  et
al. (2001) all  place the lensing galaxy in the  close vicinity of the
quasar   images,  although   to   account  for   the  relative   image
brightnesses,  this  lensing  galaxy  possesses an  implausibly  large
core. Using  a flattened potential  and explaining the  relative image
brightnesses  as due  to the  influence of  a `naked  cusp',  Lewis et
al. (2002)  find that  the lens galaxy  is offset by  $\sim0.5$ arcsec
from the quasar images.

The  data   were  examined  for   evidence  of  the   lensing  galaxy.
Unfortunately,  it was  found that  it is  not possible  to  place any
stringent   constraints    on   the   lensing    galaxy   from   these
observations. The  reason for this is  that the PSF  model we employed
does  not give  an accurate  representation of  our  observations (the
spatial profile of the 51~Peg spectra is not as strongly asymmetric as
the profile  shown in Figure~1), and  it was not  possible to subtract
off the bright quasar images to better than $\sim2\%$ using that model
PSF.  By considering the residual luminosity in dark troughs in a Keck
HIRES spectrum of  this system (Ellison et al.  1999a,b), Ibata et al.
(1999) showed  that  the  lensing   galaxy  must  be  at  least  seven
magnitudes fainter than image A, so the present constraint contributes
no additional useful information.

\section{Conclusions}\label{conclusions}
In  this paper  we have  presented spatially  resolved spectra  of the
gravitationally  lensed BAL  quasar \apm\  obtained with  STIS  on the
HST. They clearly  show that each of the  three point-like sources are
images of  the background quasar, confirming that  \apm\ represent the
first truly odd-image lens system.

An examination  of the spectra reveals significant  differences in the
equivalent widths of the Ly$\alpha + $N~V emission feature between the
images. Such  differences are naturally explained  by the differential
magnification influence  of gravitational microlensing,  which affects
the  small continuum  source but  not the  larger broad  emission line
region  of  the  QSO.    While  further  spectroscopic  monitoring  is
necessary   to  confirm   this  microlensing   hypothesis,  unresolved
ground-based observations should be adequate for this purpose (as they
will still  show variations in  the equivalent widths of  the emission
lines).

Finally,  the  spectra where  examined  for  evidence  of the  lensing
galaxy.  While the HST resolution allowed us to extract the spectra of
individual quasar  images, uncertainties in the  point spread function
of the  instrument limited  the accuracy of  the subtraction  to $\sim
2$\%. Available  imaging and spectroscopy  data show that  the lensing
galaxy must  be significantly fainter than  2\% of the  QSO images, so
that we cannot detect it in the present set of observations.


\newcommand{\mnras}{MNRAS}
\newcommand{\nat}{Nature}
\newcommand{\araa}{ARAA}
\newcommand{\aj}{AJ}
\newcommand{\apj}{ApJ}
\newcommand{\apjl}{ApJ}
\newcommand{\apjs}{ApJSupp}
\newcommand{\aap}{A\&A}
\newcommand{\aaps}{A\&ASupp}

\end{document}